\documentclass[twocolumn]{aastex631}

\shorttitle{Seasonal Insolation Variability on Early Venus}
\shortauthors{Stephen R. Kane}

\begin{document}

\title{Seasonal Insolation Variability on Early Venus: Implications
  for Energy Budget}

\author[0000-0002-7084-0529]{Stephen R. Kane}
\affiliation{Department of Earth and Planetary Sciences, University of
  California, Riverside, CA 92521, USA}
\email{skane@ucr.edu}


\begin{abstract}

Venus and Earth are similar in bulk properties yet followed
dramatically different climatic trajectories. Reconstructing the
climate evolution of Venus requires understanding how variations in
rotation rate, obliquity, orbital eccentricity, and solar luminosity
shaped the spatial and temporal distribution of incident energy and
the atmospheric response. Here we present latitude-orbital phase maps
of incident solar flux for Venus at the present epoch and at an age of
0.5~Gyr, when the Sun was fainter and Venus may have occupied a
different dynamical state. We explore endmember rotation regimes
(slow-rotator and fast-rotator), moderate obliquity (10\degr), and
elevated eccentricity ($e = 0.15$--0.30), motivated by dynamical
studies that quantify plausible limits. To translate the flux maps
into climate-relevant quantities, we apply an idealized atmospheric
energy-balance framework, including both global (0-D) and
latitude-dependent (1-D) formulations calibrated to modern Venus. This
framework is used to define a radiative relaxation timescale that
links forcing variability to expected thermal response. This approach
provides a link between orbital forcing and surface energy balance,
allowing an assessment of seasonal and orbital variability relative to
Venus's extreme greenhouse state. Our results show that, while early
Venus could experience substantial redistribution of insolation across
latitude and orbital phase, the orbit-averaged incident flux varies
only modestly across the explored parameter space and the dominant
control on surface temperature remains atmospheric opacity. Insolation
variations therefore act primarily as modulators rather than primary
drivers of climate state, with their climatic expression governed by
the competition between the forcing timescale and the radiative
adjustment time. The provided insolation maps and response diagnostics
may serve as boundary conditions for future 3-D climate simulations
that investigate the early history of Venus, including regimes in
which temperate surface conditions may have been sustained.

\end{abstract}



\section{Introduction}
\label{intro}

Venus is Earth's nearest planetary analog in size, mass, and bulk
composition, yet its present climate represents one of the most
extreme end states accessible to a terrestrial planet
\citep{kasting1988c}. The Venusian dense $\sim$93~bar CO$_2$-dominated
atmosphere produces a surface temperature near 735~K, exceeding that
expected from solar heating alone by over 500~K
\citep{limaye2018a,gillmann2022}. The divergence of Venus into its
present state can be explained by a past runaway greenhouse process,
through which surface water evaporated and was lost to space
\citep{ingersoll1969c}. Indeed, measurements of the deuterium to
hydrogen ratio (D/H) in Venus's atmosphere by \citet{donahue1982}
revealed a value about 100 times higher than Earth's D/H, implying
significant water loss over time. Understanding the precise mechanisms
of how Venus reached this state is central to comparative planetology
and to interpreting terrestrial exoplanets occupying similar
irradiation regimes
\citep{kane2014e,kane2018d,kane2019d,ostberg2019,ostberg2023a,miles2025}.

A growing body of work suggests that Venus's climatic divergence from
Earth was not inevitable, but instead depended sensitively on its
rotational and orbital evolution
\citep[e.g.,][]{way2016,krissansentotton2021c}. Slow rotation, in
particular, has been shown to promote thick dayside cloud decks that
increase planetary albedo and reduce surface temperatures under
otherwise intense stellar forcing \citep{way2020}. Alternatively,
water condensation may have never have occurred on the Venusian
surface, resulting in a rapid transition to its present state
\citep{hamano2013,turbet2021,constantinou2025}. Furthermore, changes
in obliquity or orbital eccentricity can redistribute incident flux
both spatially and temporally, potentially triggering feedbacks that
accelerate atmospheric loss or greenhouse runaway
\citep{williams2003a,atobe2007,spiegel2009a,kane2012e,barnes2013a,armstrong2014b,linsenmeier2015,barnes2016a,kane2020e,vervoort2022,way2023b}. Today,
Venus has a very slow retrograde rotation, nearly zero obliquity, and
an almost circular orbit, resulting in minimal seasonal variation in
insolation. While modern 3-D general circulation models (GCMs) are
indispensable for capturing the full climate dynamics of Venus, there
remains a need for physically transparent, intermediate-level
approaches that isolate the role of orbital and rotational forcing
\citep{cowan2012c}. Latitude-orbital phase maps of incident stellar
flux provide such a bridge, offering direct insight into the seasonal
energy distribution that ultimately drives atmospheric response
\citep{dobrovolskis2013b,kane2017d}.

In this paper, we extend these techniques to Venus, focusing on
conditions at $\sim$0.5~Gyr after formation. At this epoch, the Sun
was significantly fainter, and Venus may have occupied dynamical
states substantially different from those observed today. By combining
flux maps with an idealized greenhouse energy balance model and
radiative equilibrium timescales, we quantify the extent to which
seasonal and orbital variations could have influenced Venus's energy
budget during its early evolution. In Section~\ref{dynamics}, we
describe the dynamical evolution of Venus and the extent to which
these properties may have varied in the past. Section~\ref{flux}
presents the methodology and outcome of calculating longitude-orbital
phase flux variations for Venus for various dynamical values at an age
of $\sim$0.5~Gyr, and in comparison to the present
epoch. Section~\ref{balance} translates the flux map calculations into
models of planetary temperatures and radiative response times. We
discuss the implications of these results in Section~\ref{discussion},
and provide concluding remarks in Section~\ref{conclusions}.


\section{Venusian Dynamical Evolution}
\label{dynamics}

\begin{figure*}
  \begin{center}
    \includegraphics[width=16.0cm]{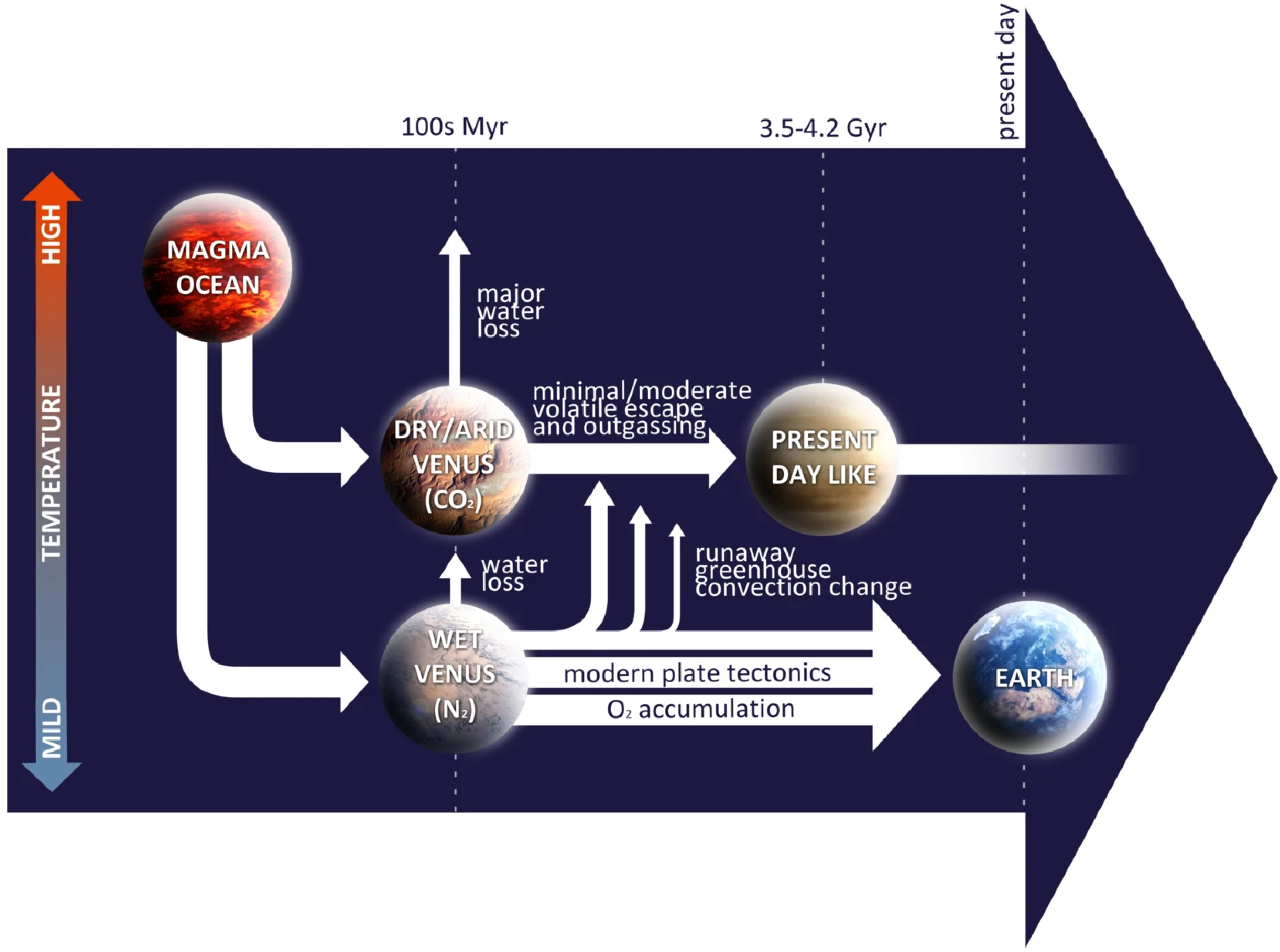}
  \end{center}
  \caption{Representation of possible evolutionary pathways for Venus
    to its present state, starting with a magma ocean phase. The top
    pathway is for the case where Venus lost much of its water
    inventory early, while the bottom pathway considers an extended
    period of surface liquid water until the climate was
    destabilized. Figure reproduced from \citet{gillmann2022}.}
  \label{fig:gillmann}
\end{figure*}

As described in Section~\ref{intro}, there are currently multiple
viable evolutionary pathway scenarios for Venus that are broadly
consistent with the available data and models. Shown in
Figure~\ref{fig:gillmann} is a representation for several of these
pathways \citep{gillmann2022}. The top ``dry Venus''pathway assumes
that an extended magma ocean phase combined with the atmospheric
energy balance may have prevented surface water condensation
\citep{hamano2013,salvador2023b}. Under this scenario, the water vapor
may have formed clouds on the night side of the planet
\citep{turbet2021}, eventually resulting in the loss of the water
inventory through early intense hydrodynamic escape
\citep{gillmann2009b}. Alternatively, the bottom pathway accounts for
the possible formation of surface water oceans on Venus, enabling
clement conditions for as long as several Gyr depending on rotation
rate \citep{yang2014b}, before the planet succumbed to runaway
greenhouse mechanisms \citep{way2016,way2020,kane2024b}. Regardless,
it is clear that the evolution of Venus into its present state is a
sensitive function of the initial conditions, including formation and
composition, as well as the dynamical and insolation flux components
and the changes in those components that occurred thereafter.

The rotational and orbital evolution of Venus is governed by a complex
interplay between gravitational tides, atmospheric thermal tides,
core-mantle coupling, and planetary perturbations
\citep{takagi2007,cottereau2011,limaye2018a,rolf2022,widemann2023}. Early
studies demonstrated that Venus possesses multiple stable spin
equilibria arising from the competition between solid-body tides and
atmospheric thermal tides \citep{correia2003a,correia2003b}. These
equilibria include both prograde and retrograde states with rotation
periods ranging from tens to hundreds of Earth days. Importantly,
capture into any one of these states depends on initial rotation rate,
obliquity, and atmospheric structure. More recent work has
incorporated evolving stellar luminosity and realistic atmospheric
torques. For example, \citet{revol2023} showed that, when solar
luminosity evolution is included, Venus-like planets may never reach a
strict rotational equilibrium, instead evolving continuously as
atmospheric tides weaken over time. \citet{musseau2024} further
demonstrated that solid-body tidal dissipation alone cannot account
for the present rotation of Venus if the planet began with a very
rapid spin, reinforcing the importance of atmospheric processes
\citep{margot2021c,kane2022b,revol2023}.

Obliquity evolution is similarly constrained. Unlike Earth, Venus
lacks a large satellite capable of stabilizing its obliquity, making
its spin-state particularly sensitive to dynamical forcing
\citep{laskar1993a,laskar1993b}. Numerical integrations indicate that
early Venus likely maintained low obliquity, with variations of only a
few degrees over Gyr timescales
\citep{barnes2016a}. Nevertheless, transient excursions to obliquities
of order 5--10\degr are dynamically plausible and can introduce
meaningful seasonal forcing.

Long-term dynamical studies indicate that Venus's orbital eccentricity
may have undergone a range of evolutionary pathways, depending on
formation/migration scenarios and initial conditions. Ensemble
integrations of the secular Solar System over 5~Gyr show that the
inner planets undergo chaotic diffusion, with Venus's eccentricity
probability density concentrated at low values ($\sim$0.01)
\citep{laskar2008,laskar2009c}. Further simulations of the late stages
of terrestrial planet accretion typically yield Earth/Venus analogs
with eccentricities larger than the present terrestrial values
\citep{chambers2001c}, which are eventually damped via dynamical
friction \citep{obrien2006b}. An alternative scenario that produces an
early relatively high Venus eccentricity was proposed by
\citet{kane2020e}, who showed that plausible shifts in Jupiter's orbit
may have excited Venus's eccentricity as high as $e\simeq0.3$,
followed by damping back to the current near-circular orbit through
tidal dissipation provided by a substantial primordial water
inventory. Such eccentricities would produce strong perihelion heating
episodes, potentially influencing atmospheric escape and climate
stability.

Motivated by these various previous studies, we adopt a set of
``maximum plausible'' dynamical parameters at at age of 0.5~Gyr. For
the rotation period, two states of either a slow rotator
(approximately tidally locked, as is its present state) or a rapid
rotator ($\leq 5$~days). For the obliquity, we consider values as high
as 10\degr, compared with its present value of 2.64\degr. For the
orbital eccentricity, we model values as high as $e = 0.15$, compared
with its present value of $e = 0.007$; the lowest of all Solar System
planets at the present epoch.


\section{Incident Flux Maps}
\label{flux}


\subsection{Flux Map Formalism}
\label{form}

To compute top-of-atmosphere (TOA) incident stellar flux, we adopt the
methodology described by \citet{kane2017d}. This formalism
parameterizes the flux in terms of the latitude and orbital phase of
the planet, accounting for the combined effects of orbital geometry,
obliquity, eccentricity, and the rotation regime. For a planet
orbiting a star of luminosity $L_\star$, the instantaneous stellar
flux received at distance $r$ is
\begin{equation}
  F_p = \frac{L_\star}{4\pi r^2}.
  \label{eqn:flux}
\end{equation}
The orbital distance varies with true anomaly $f$ as
\begin{equation}
  r = \frac{a(1-e^2)}{1 + e \cos f},
\end{equation}
where $a$ is the semi-major axis and $e$ is the eccentricity. The
maximum flux received at a given latitude, $\beta$, is given by
\begin{equation}
  F_p = \frac{L_\star}{4 \pi r^2} \cos | \beta - \delta |
  \label{eqn:slow}
\end{equation}
The solar declination, $\delta$, is given by
\begin{equation}
  \delta = \varepsilon \cos [2 \pi (\phi - \Delta \phi)]
\end{equation}
for which $\phi$ is the orbital phase, $\Delta \phi$ is the offset in
phase between periastron and highest solar declination in the northern
hemisphere, and $\varepsilon$ is the obliquity.

Equations~\ref{eqn:slow}--\ref{eqn:fast} below describe two limiting
interpretations of the received flux. For slowly rotating planets,
where the rotation period is comparable to the orbital period, the
sub-stellar point remains nearly fixed in longitude over long time
intervals. In this limit, Equation~\ref{eqn:slow} gives the maximum
(sub-stellar) flux at a given latitude and orbital phase; other
longitudes receive less flux, so the slow-rotator maps presented in
subsequent sections represent the upper envelope of local forcing at
each latitude rather than a longitudinally uniform quantity. For
rapidly rotating planets, where the rotation rate is $\ll$ the orbital
period, longitudinal contrasts are efficiently smoothed by diurnal
cycling. Averaging Equation~\ref{eqn:slow} over all hour angles yields
a flux that depends only on latitude and orbital phase
\citep{berger1978c}; the resulting fast-rotator maps therefore
represent a true diurnal-mean quantity applicable at all
longitudes. In the fast-rotator case, the flux expression becomes
\begin{equation}
  F = \frac{L_\star}{4 \pi r^2} (\sin \delta \sin \beta + \cos \delta
  \cos \beta \cos h)
  \label{eqn:fast}
\end{equation}
where $h$ is the hour angle of the star with respect to the local
meridian. The fraction of planetary rotation period that experiences
daylight for a given latitude is
\begin{equation}
  \Delta t_{dl} = \frac{2 \arccos(-\tan \delta \tan \beta)}{360\degr}
  \label{daylight}
\end{equation}
Note that, for obliquities of $\varepsilon > 0\degr$, there exist
latitudes of the planet that experience constant day/night during the
course of an orbital period.


\subsection{Current Epoch}
\label{current}

\begin{figure*}
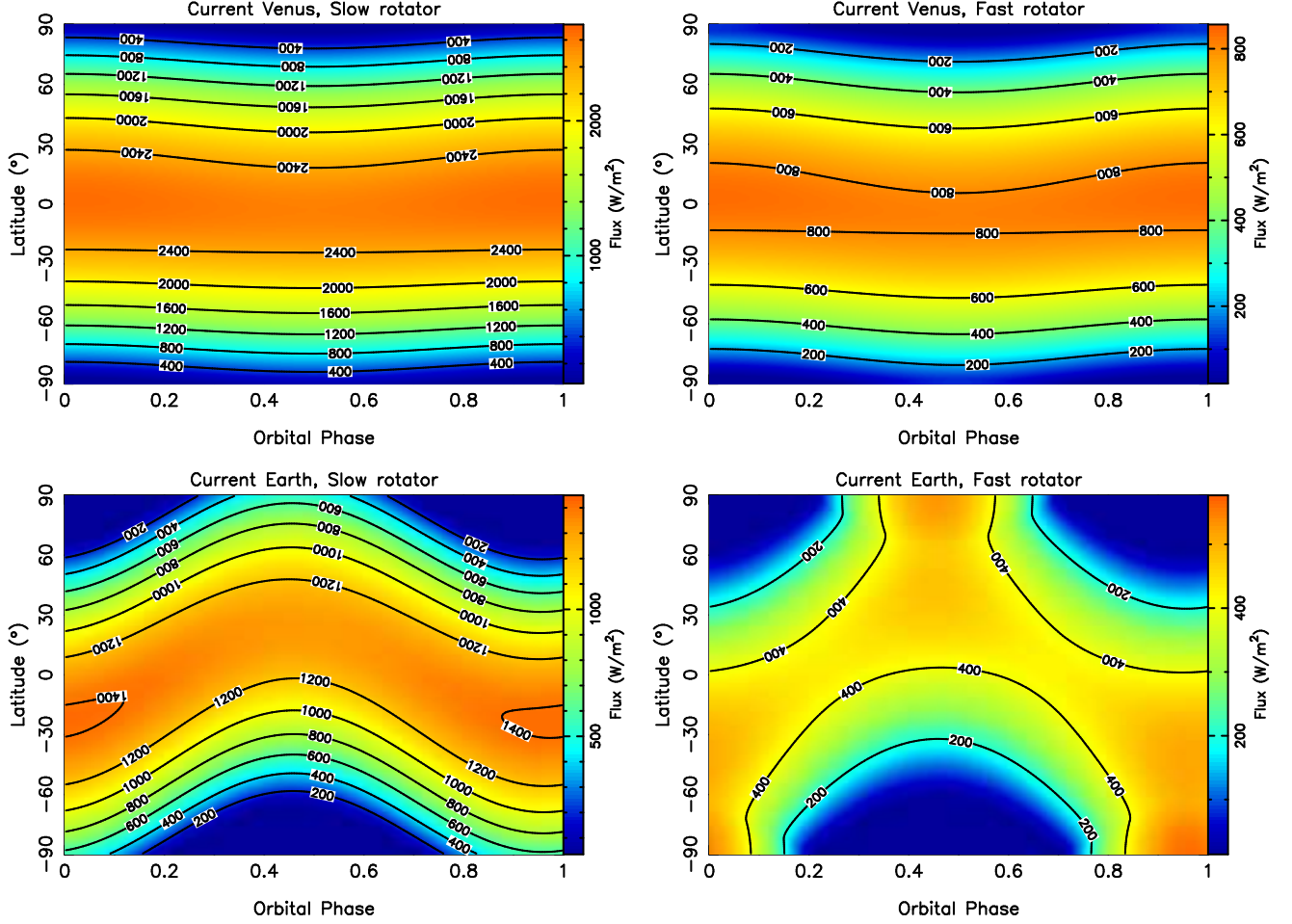

    \begin{center}
        \begin{tabular}{cc}
            \includegraphics[angle=270,width=8.5cm]{f02a.eps} &
            \includegraphics[angle=270,width=8.5cm]{f02b.eps} \\
            \includegraphics[angle=270,width=8.5cm]{f02c.eps} &
            \includegraphics[angle=270,width=8.5cm]{f02d.eps} \\
        \end{tabular}
    \end{center}
  \caption{TOA flux maps for Venus (top row) and Earth (bottom row) as
    a function of planetary latitude and orbital phase at the present
    epoch of solar luminosity, and with contours of constant flux
    (W/m$^2$) overlaid. The flux maps are represented for the
    slow-rotator (left column) and fast-rotator (right column)
    scenarios. Note that the slow-rotator maps show the maximum
    (sub-stellar) flux at each latitude, whereas the fast-rotator maps
    show the diurnally averaged flux, which is applicable at all
    longitudes (see Section~\ref{form}). The two representations are
    therefore not directly comparable in absolute magnitude; they
    instead bracket the range of local forcing regimes.}
  \label{fig:current}
\end{figure*}

As a baseline for the early-Venus experiments presented in the
subsequent subsections, we first compute the TOA incident flux maps
for Venus and Earth using their present-day orbital and spin-axis
configurations
\citep{tomasko1980d,tremberth2009b,kopp2011a,kopp2023}. Figure~\ref{fig:current}
shows the results of these calculations as a function of planetary
latitude and normalized orbital phase, where an orbital phase of zero
corresponds to perihelion passage \citep{laskar2004c}. The flux maps
for Venus and Earth are shown in the top and bottom rows,
respectively. The statistics (maximum, mean, median, and standard
deviation) for these flux maps are shown in the top two rows of
Table~\ref{tab:stats}. Following the flux-map formalism implemented in
the same code framework used for \citet{kane2017d}, we present two
limiting rotational states (see Section~\ref{form}). For the
slow-rotator case (left column), the local forcing retains strong
diurnal structure (i.e., the instantaneous or near-instantaneous
geometry dominates), producing higher peak fluxes and sharper spatial
gradients that emphasize day-side illumination. In the fast-rotator
case (right column), diurnal averaging strongly reduces the
instantaneous peaks and yields a smoother latitudinal distribution
that more closely reflects the seasonal evolution of the subsolar
latitude and the latitude-dependent day length
\citep{berger1978c}. This paired representation is useful because it
isolates the role of rotation in translating orbital geometry into an
effective climatic forcing, while keeping the orbital configuration
fixed.

The present-epoch comparison also highlights the fundamentally
different nature of seasonal forcing on Venus versus Earth. Earth's
substantial obliquity produces a clear seasonal migration of peak
insolation between hemispheres, with high-latitude maxima during
summer seasons and a pronounced temporal modulation over the orbit
(Figure~\ref{fig:current}, bottom-right). In contrast, Venus's
contemporary configuration yields comparatively weak seasonal
redistribution with the latitudinal structure close to symmetric about
the equator and varies only modestly with orbital phase
(Figure~\ref{fig:current}, top-left). The primary difference between
the Venus slow and fast rotator panels is therefore the amplitude and
sharpness of the forcing rather than the timing of seasonal extremes
\citep{tomasko1980d,haus2016,limaye2018a}.


\subsection{Solar Luminosity Evolution}
\label{solar}

The Habitable Zone (HZ) is defined as the region around a star where a
planet may have surface liquid water given sufficient atmospheric
pressure, the extent of which is governed by the effective temperature
and luminosity of the host star
\citep{kasting1993a,kane2012a,kopparapu2013a,kopparapu2014,chandler2016,hill2023}. The
HZ may be further subdivided into the conservative HZ (CHZ), defined
by the runaway and maximum greenhouse limits, and the optimistic HZ
(OHZ), defined by empirically derived estimates of possible surface
liquid water on Venus and Mars \citep{kasting1993a,kane2016c}. The
solar luminosity is known to have increased monotonically since the
zero-age main sequence \citep{gough1981d}. For example, at an age of
0.5~Gyr, the solar luminosity was approximately 73\% of its present
value.

\begin{deluxetable*}{lrrr|rrrr|rrrr} \label{tab:stats}
    \tablecaption{Flux map statistics, where all flux units are W/m$^2$. The statistics (maximum, mean, median, and standard deviation) are computed over all latitude--orbital-phase grid cells in each map. For the slow-rotator columns, the values refer to sub-stellar (maximum) flux at each latitude, whereas for the fast-rotator columns they refer to diurnally averaged flux (see Section~\ref{form}). The two sets of statistics are therefore not directly comparable in absolute terms.}
    \tablecolumns{12}
    \tablewidth{0pt}
    \tablehead{
        \multicolumn{4}{c}{} &
        \multicolumn{4}{c}{Slow-rotator} &
        \multicolumn{4}{c}{Fast-rotator} \\
        \colhead{Planet} &
        \colhead{Age} &
        \colhead{$\varepsilon$ (\degr)} &
        \colhead{$e$} &
        \colhead{Max} &
        \colhead{Mean} &
        \colhead{Median} &
        \colhead{$\sigma$} &
        \colhead{Max} &
        \colhead{Mean} &
        \colhead{Median} &
        \colhead{$\sigma$}
    }
    \startdata
Venus & 4.6 &  2.64 & 0.007 & 2648.0 & 1661.9 & 1845.5 &  804.8 &  835.7 &  525.0 &  583.0 &  252.7 \\
Earth & 4.6 & 23.44 & 0.017 & 1413.7 &  851.6 &  965.1 &  450.4 &  562.8 &  297.0 &  349.2 &  165.2 \\
\hline
Venus & 0.5 &  2.64 & 0.007 & 1937.8 & 1216.2 & 1350.6 &  588.9 &  611.6 &  384.2 &  426.6 &  184.9 \\
Venus & 0.5 & 10.00 & 0.007 & 1937.9 & 1211.9 & 1350.5 &  597.0 &  624.8 &  391.7 &  434.6 &  188.3 \\
Venus & 0.5 &  2.64 & 0.150 & 2644.7 & 1230.1 & 1279.8 &  664.2 &  834.6 &  388.7 &  403.9 &  208.8 \\
Venus & 0.5 & 10.00 & 0.150 & 2644.8 & 1225.7 & 1279.7 &  671.6 &  852.7 &  396.5 &  415.9 &  213.6 \\
    \enddata
\end{deluxetable*}

\begin{figure*}
  \begin{center}
    \includegraphics[angle=270,width=16.0cm]{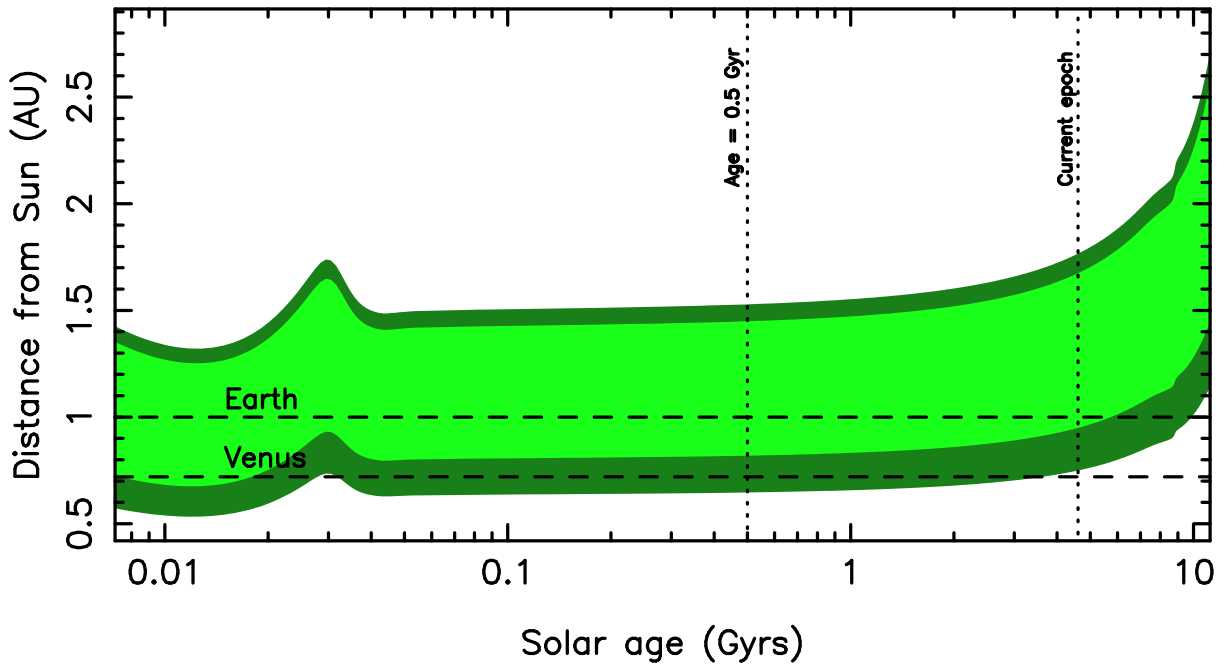}
  \end{center}
  \caption{The evolution of the Solar System HZ as a function of solar
    age. The CHZ and OHZ are represented by the light and dark green
    regions, respectively. The locations of Venus and Earth are
    represented by horizontal dashed lines, and the vertical dotted
    lines indicate the 0.5~Gyr epoch and the current epoch.}
  \label{fig:hzevolve}
\end{figure*}

To simulate the evolution of the solar properties and their influence
on the insolation flux and HZ, we utilized the BaSTI Stellar Evolution
Models and Isochrones \citep{hidalgo2018}. The solar evolution track
was produced through the selection of a solar mass star with solar
initial metallicity and a helium mass fraction of $Z_\odot = 0.01721$,
$Y_\odot = 0.2695$. The luminosity and effective temperature values
from this evolutionary track were used to calculate the HZ boundaries
through time, and are depicted in Figure~\ref{fig:hzevolve}. The CHZ
and OHZ are represented by the light and dark green regions,
respectively. The time range includes the pre-main sequence period,
and extends up to the transition of the Sun into the sub-giant
branch. The horizontal dashed lines indicate the semi-major axes of
Venus and Earth, and the vertical dotted lines represent the 0.5~Gyr
epoch and the current epoch. The relative insolation fluxes received
by Venus and Earth at the current epoch are 1.91 and 1.0,
respectively. At an age of 0.5~Gyr, those relative fluxes are 1.41 and
0.73, respectively, when the Sun was 73\% of its present value,
Resulting in a significantly reduced flux distribution over the
surface of the terrestrial planets.

\subsection{Maximum Rotation}
\label{rot}

\begin{figure*}
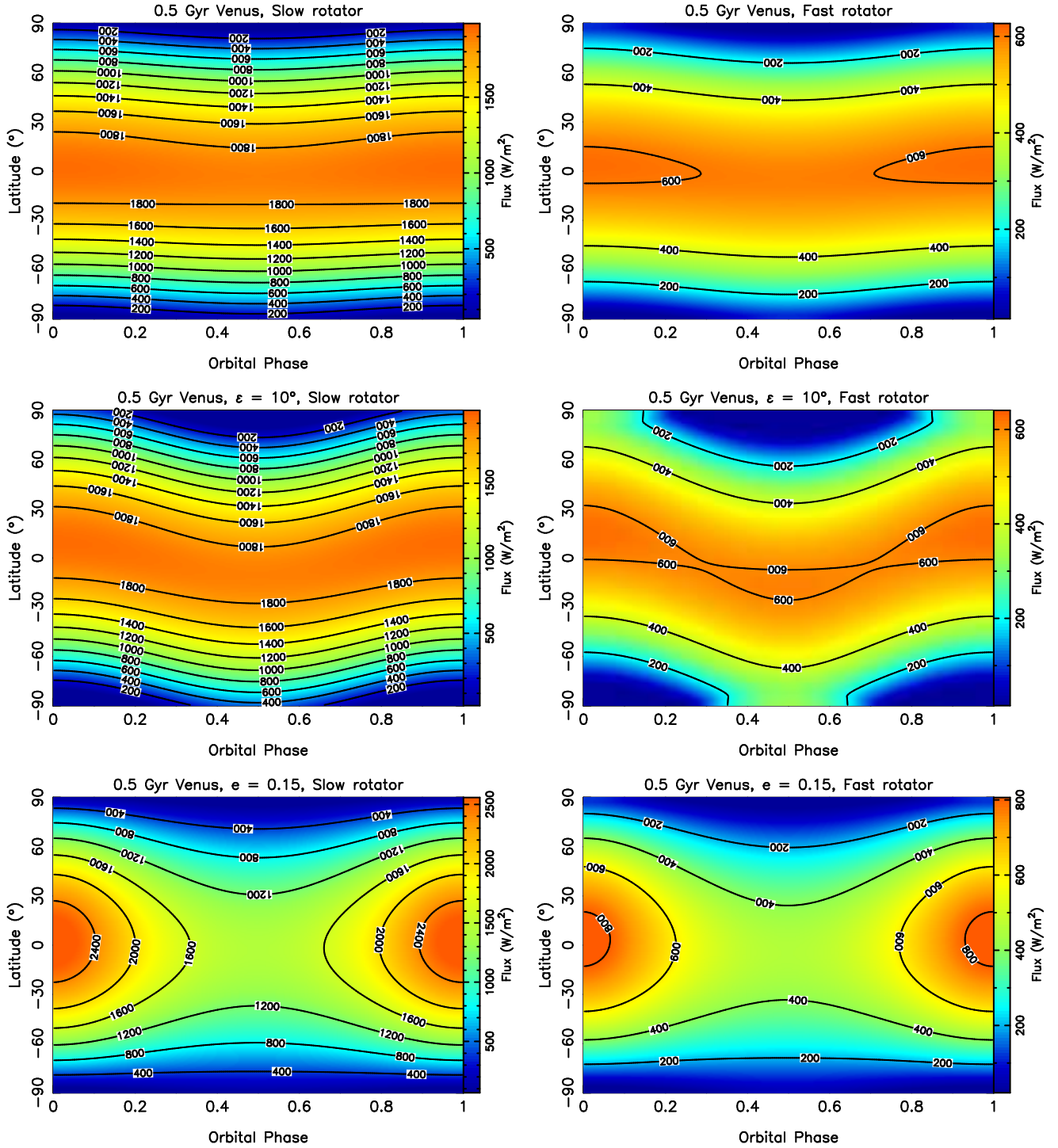

    \begin{center}
        \begin{tabular}{cc}
            \includegraphics[angle=270,width=8.5cm]{f04a.eps} &
            \includegraphics[angle=270,width=8.5cm]{f04b.eps} \\
            \includegraphics[angle=270,width=8.5cm]{f04c.eps} &
            \includegraphics[angle=270,width=8.5cm]{f04d.eps} \\
            \includegraphics[angle=270,width=8.5cm]{f04e.eps} &
            \includegraphics[angle=270,width=8.5cm]{f04f.eps} \\
        \end{tabular}
    \end{center}
  \caption{TOA flux maps for Venus at an age of 0.5~Gyr using
    $\varepsilon = 2.64\degr$ and $e = 0.007$ (top row,
    Section~\ref{rot}), $\varepsilon = 10.0\degr$ and $e = 0.007$
    (middle row, Section~\ref{obl}), and $\varepsilon = 2.64\degr$ and
    $e = 0.15$ (bottom row, Section~\ref{ecc}). The flux maps are
    plotted as a function of planetary latitude and orbital phase at
    73\% of the present solar luminosity, and with contours of
    constant flux (W/m$^2$) overlaid. The flux maps are represented
    for the slow-rotator (left column) and fast-rotator (right column)
    scenarios. As in Figure~\ref{fig:current}, the slow-rotator panels
    show sub-stellar (maximum) flux, while the fast-rotator panels
    show the diurnal average.}
  \label{fig:mod}
\end{figure*}

Motivated by the range of spin states described in
Section~\ref{dynamics}, we adopt two limiting rotation regimes for
early Venus at 0.5~Gyr of a slow (representative of synchronous or
near-synchronous rotation) and fast (diurnally averaged) rotational
states (see Section~\ref{form}). Such endmember states are well
supported by dynamical studies demonstrating multiple stable (or
long-lived) rotation outcomes for Venus-like planets, including both
prograde and retrograde solutions and a broad range of rotation
periods depending on atmospheric structure and tidal dissipation
efficiency \citep{correia2003a,correia2003b,revol2023,musseau2024}. We
also adopted the current obliquity ($\varepsilon = 2.64\degr$) and
eccentricity ($e = 0.007$) values for Venus. The resulting TOA flux
maps are shown in the top row of Figure~\ref{fig:mod}, where the
slow-rotator case retains the strong instantaneous geometry of
Equation~\ref{eqn:slow}, and the fast-rotator case reflects the
diurnally averaged forcing captured by Equation~\ref{eqn:fast} and the
daylight fraction (Equation~\ref{daylight}).

At these fixed orbital elements, the dominant impact of changing the
rotation regime is a redistribution of the forcing from a sharply
peaked instantaneous pattern to a smoother, latitude-dependent
seasonal pattern. This is reflected by the summary statistics in
Table~\ref{tab:stats} (row 3), which shows that the slow-rotator map
yields a substantially larger peak flux ($\sim$1938~W/m$^2$) than the
fast-rotator map ($\sim$612~W/m$^2$), as expected when comparing
instantaneous (subsolar) forcing to diurnally averaged flux
\citep{berger1978c}. The diurnal averaging also reduces the overall
spread of forcing across latitude and phase, with the standard
deviation decreasing from $\sigma \simeq 589$~W/m$^2$ (slow-rotator)
to $\sigma \simeq 185$~W/m$^2$ (fast-rotator). These rotational
endmembers therefore bracket the range of TOA forcing patterns that
could have governed early Venus climate responses, including the
strength of day-night contrasts and the degree to which seasonal
variability is expressed in the local energy input
\citep{barnes2016a,way2016}.


\subsection{Maximum Obliquity}
\label{obl}

Although Venus today has a near-zero obliquity, numerical integrations
show that early Venus could plausibly have experienced modest
obliquity excursions, with values of order several degrees to
$\sim$10\degr, depending on the assumed spin-state history and
perturbations \citep{laskar1993a,laskar1993b,barnes2016a}. To quantify
the impact of such excursions on the spatial and temporal distribution
of insolation, we adopt an obliquity of $\varepsilon = 10\degr$ at
0.5~Gyr and compute flux maps for both rotational states. The
corresponding maps are shown in the middle row of
Figure~\ref{fig:mod}. Relative to the baseline 0.5~Gyr case (top row),
the elevated obliquity produces an orbital-phase-dependent migration
of the subsolar latitude (via the declination term in
Section~\ref{form}), as expected. This manifests as a larger seasonal
modulation at mid-to-high latitudes and the appearance of more
pronounced phase-dependent asymmetries between hemispheres.

Despite these clear geometric differences, the global statistics of
the forcing are only weakly modified for this modest obliquity change,
consistent with the expectation that obliquity redistributes flux in
latitude and season without altering the global-mean incident
energy. Table~\ref{tab:stats} (row 4) shows that the slow-rotator
maximum, mean, and median fluxes are essentially unchanged compared to
the baseline case (row 3), while the standard deviation increases
slightly, reflecting enhanced seasonal contrast at higher
latitudes. In the fast-rotator case, the same behavior is evident,
accompanying the more pronounced seasonal latitudinal redistribution
(Figure~\ref{fig:mod}). These results emphasize that even relatively
small obliquities can introduce measurable seasonal forcing in the TOA
flux geometry, providing an important control parameter for subsequent
energy-balance interpretation and for comparisons with Venus climate
states that may be sensitive to the latitude-season structure of
insolation \citep{williams2003a,spiegel2009a,barnes2016a}.


\subsection{Maximum Eccentricity}
\label{ecc}

In addition to spin-axis evolution, early Venus may have experienced
significantly elevated orbital eccentricity compared to its present
nearly circular orbit \citep{chambers2001c,kane2020e}, possibly
resulting in strong perihelion heating episodes and enhanced seasonal
modulation of the incident flux. Here we quantify this effect by
adopting $e = 0.15$ at 0.5~Gyr (with $\varepsilon=2.64\degr$) and
computing TOA flux maps for the slow and fast rotational regimes. The
resulting maps are shown in the bottom row of
Figure~\ref{fig:mod}. Since the orbital phase is defined such that
$\phi = 0$ corresponds to perihelion passage (Section~\ref{current}),
eccentricity-driven forcing appears as a pronounced enhancement of
flux near $\phi \simeq 0$ and a corresponding reduction near aphelion,
with the amplitude set primarily by the $r^{-2}$ dependence of
Equation~\ref{eqn:slow} and Equation~\ref{eqn:fast}.

The flux statistics in Table~\ref{tab:stats} (row 5) reflect this
strongly non-uniform distribution in orbital phase. The slow-rotator
maximum increases from 1938~W/m$^2$ in the low-eccentricity baseline
case to $\sim$2645~W/m$^2$ for $e = 0.15$, consistent with the
analytic perihelion scaling $F_p \propto (1-e)^{-2}$ at fixed
semi-major axis. The variance also increases appreciably ($\sigma
\simeq 664$~W/m$^2$), while the median decreases to
$\sim$1280~W/m$^2$), indicating that the flux map is increasingly
dominated by brief, high-flux intervals near perihelion rather than a
uniform enhancement across the orbit. In the fast-rotator case, the
same qualitative behavior occurs: the maximum rises to
$\sim$835~W/m$^2$ and the spread increases ($\sigma \simeq
209$~W/m$^2$), while the mean and median values undergo only modest
changes, as expected. These eccentricity-driven excursions therefore
act primarily as intermittent climate forcing pulses that may be
important when atmospheric radiative or dynamical adjustment
timescales are shorter than, or comparable to, the duration of
perihelion forcing \citep{atobe2007,kane2012e,kane2020e}.


\subsection{Combining All Factors}
\label{comb}

\begin{figure*}
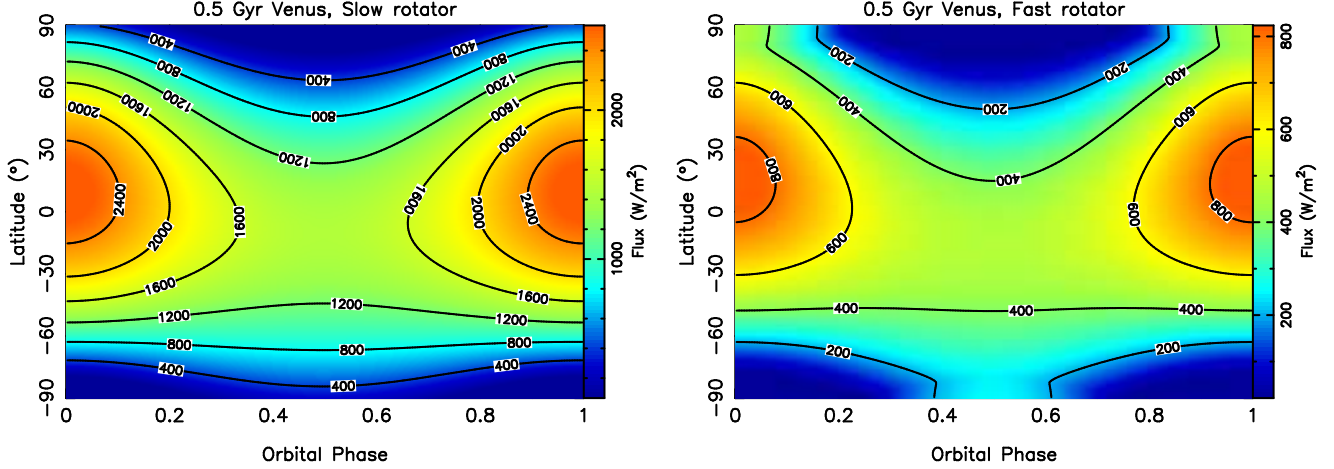

    \begin{center}
        \begin{tabular}{cc}
            \includegraphics[angle=270,width=8.5cm]{f05a.eps} &
            \includegraphics[angle=270,width=8.5cm]{f05b.eps} \\
        \end{tabular}
    \end{center}
  \caption{TOA flux maps for Venus at an age of 0.5~Gyr using the
    combination of $\varepsilon = 10.0\degr$ and $e = 0.15$ (see
    Section~\ref{comb}). The flux maps are plotted as a function of
    planetary latitude and orbital phase at 73\% of the present solar
    luminosity, and with contours of constant flux (W/m$^2$)
    overlaid. The flux maps are represented for the slow-rotator
    (left) and fast-rotator (right) scenarios. The slow-rotator panel
    shows sub-stellar (maximum) flux, while the fast-rotator panel
    shows the diurnal average.}
  \label{fig:comb}
\end{figure*}

The final scenario combines the maximum (plausible) dynamical
contributors to seasonal and annual insolation variability at the
0.5~Gyr epoch by adopting an obliquity of $\varepsilon = 10\degr$ and
an eccentricity of $e = 0.15$, and evaluating both slow and fast
rotational limits, thus representing an upper envelope of TOA forcing
variability. The resulting flux maps, shown in Figure~\ref{fig:comb},
exhibit the superposition of eccentricity-driven flux enhancements
concentrated near perihelion and obliquity-driven migration of the
subsolar latitude through the orbital phase dependence of the solar
declination. As in the preceding subsections, the rotational regime
determines how strongly the instantaneous geometry is expressed in the
forcing. The slow-rotator case preserves sharper spatial gradients and
larger instantaneous peaks (left panel), whereas the fast-rotator case
produces a smoother latitudinal structure through diurnal averaging
(right panel). Together, these features provide a physically intuitive
forcing template for assessing how an early Venus atmosphere might
integrate or respond to short-duration perihelion flux pulses
superimposed on seasonal latitudinal redistribution. We note that the
combined scenario introduces an additional degree of freedom not
explored here: the longitude of perihelion relative to the solstice
(i.e., the phase offset $\Delta\phi$ in Section~\ref{form}). This
parameter controls whether perihelion coincides with maximum or
minimum solar declination in a given hemisphere, and can substantially
alter the seasonal asymmetry between hemispheres
\citep{spiegel2010b,linsenmeier2015,ohno2019a,ohno2019b,guendelman2020}. A
full exploration of this parameter space is deferred to future work.

The combined-factor statistics are summarized in Table~\ref{tab:stats}
(row 6) and demonstrate that, for the slow-rotator regime, the maximum
flux ($\sim$2645~W/m$^2$) remains essentially set by the
eccentricity-enhanced perihelion forcing, while the overall spread
increases slightly relative to the maximum eccentricity case (row
5). In the fast-rotator regime, the combined case yields a modestly
larger peak ($\sim$853~W/m$^2$) and variance than the maximum
eccentricity case, reflecting the additional seasonal redistribution
introduced by the non-zero obliquity. Notably, the mean and median
values remain close to those of the other 0.5~Gyr scenarios,
reinforcing that obliquity and eccentricity primarily redistribute
flux in latitude and phase rather than strongly altering the
orbit-averaged energy input. This behavior is particularly relevant
for the atmospheric energy-balance. If atmospheric radiative and
dynamical response times are long compared to the orbital timescale,
the climate will tend to track the mean forcing, whereas shorter
response times may allow partial adjustment to perihelion-driven
transients and their hemispheric seasonal modulation
\citep{atobe2007,kane2012e,kane2020e}.

\section{Atmospheric Energy Balance}
\label{balance}

The flux maps presented in Section~\ref{flux} quantify how the
incident solar energy is distributed over latitude and orbital phase
for a range of rotational regimes, obliquities, and eccentricities. To
connect these geometric forcing patterns to climate-relevant behavior,
we require a framework that (i) provides an interpretable mapping
between absorbed shortwave flux and radiated longwave emission, and
(ii) quantifies the characteristic time over which an atmosphere can
thermally adjust to changes in forcing. In this section we adopt a
simple idealized greenhouse model for the mean state, and a radiative
relaxation timescale, $\tau_E$, for the thermal response. A key point
is that $\tau_E$ is primarily a property of the atmospheric column
(mass, heat capacity, emissivity, and radiating temperature) rather
than of the orbital geometry itself
\citep{cronin2013b,guendelman2019}. Rotation, obliquity, and
eccentricity enter through the structure and frequency content of the
forcing (i.e., the amplitude, spatial pattern, and timescale of
$F(\beta,\phi)$), which determines whether the atmosphere can track
the variations or instead smooth them into an approximately steady
mean state \citep{donohoe2020a}.


\subsection{Idealized Greenhouse Framework and Calibration}
\label{igm}

We first define a baseline (global-mean) radiative equilibrium
temperature using the expression for incident flux provided by
Equation~\ref{eqn:flux} to produce
\begin{equation}
  T_{eq} = \left( \frac{F_p (1-A_B)}{4 \sigma}
  \right)^\frac{1}{4}
  \label{eqn:teqflux}
\end{equation}
where $A_B$ is the planetary Bond albedo. Equation~\ref{eqn:teqflux}
reproduces the Earth effective radiating temperature when evaluated
with present-day parameters and serves as a convenient reference
scale. To incorporate greenhouse warming, we employ
the idealized single-layer greenhouse model in which an atmospheric
layer of temperature $T_a$ absorbs and emits longwave radiation with
effective emissivity $\epsilon$ (assumed gray in the thermal
infrared). In this case the surface temperature becomes
\begin{equation}
  T_s = \left( \frac{F_p (1-A_B)}{4 \sigma} \frac{1}{1 - \epsilon/2}
  \right)^\frac{1}{4},
  \label{eqn:teqigm}
\end{equation}
with $T_s = 2^{1/4}T_a$ for the single-layer configuration. This
parameterization is intentionally idealized in that it neglects
vertical structure, non-gray spectral windows, clouds as an explicit
radiative agent, and latent heat transport. Nevertheless, it provides
the needed mapping between absorbed shortwave energy and an effective
surface thermal state, and it can be tuned (via $A_B$ and $\epsilon$)
to approximate modern conditions. For Venus, observationally
constrained radiative energy-balance studies indicate that the
planet's effective emission occurs near the cloud tops and that the
present-day radiative environment is strongly shaped by clouds and
aerosol absorption/scattering
\citep{tomasko1980d,haus2016,limaye2018a}. In the context of this
work, we therefore treat $A_B$ and $\epsilon$ as representing an
``effective'' radiating layer and use the model primarily as a
diagnostic tool to interpret how changes in the distribution of
incident flux may translate into changes in radiative equilibrium and
response.


\subsection{Radiative Relaxation Timescale}
\label{tau}

A complementary quantity to the mean-state temperature is the
radiative equilibrium (or radiative relaxation) timescale, $\tau_E$,
which characterizes how rapidly an atmospheric column can thermally
adjust to a perturbation in radiative forcing
\citep{cronin2013b,guendelman2019,kane2021e}. This timescale controls
whether the atmosphere tracks time-variable forcing (strong
seasonal/annual response) or instead smooths variability into a nearly
steady state. For vertically extended atmospheres, the radiative
timescale is itself a function of pressure level, with deeper layers
generally exhibiting longer response times \citep{showman2002}. The
radiative relaxation timescale may be expressed as
\begin{equation}
  \tau_E = \frac{P_s c_P}{4 (2-\epsilon) \epsilon \sigma g T_{ao}^3}.
  \label{eqn:ret}
\end{equation}
where $c_P$ is the heat capacity at constant pressure, $P_s$ is the
pressure representing the effective mass of the thermally responding
column, $T_{ao}$ is the radiative equilibrium temperature, and $g$ is
surface gravity. For atmospheres with strong vertical stratification,
$P_s$ may be interpreted as an effective pressure depth participating
in the response on the timescale of interest rather than the full
surface pressure.

Equation~\ref{eqn:ret} shows that $\tau_E$ is strongly temperature
dependent ($\propto T_{ao}^{-3}$) and increases with atmospheric
column mass and heat capacity. As previously stated, for fixed
atmospheric properties ($P_s$, $c_P$, $\epsilon$), $\tau_E$ does not
explicitly depend on rotation, obliquity, or eccentricity. The
geometric parameters instead enter by determining the forcing
$F(\beta,\phi)$ and thus the relevant equilibrium temperature scales
and forcing frequencies to which the atmosphere is responding.


\subsection{Connecting Flux Maps to Thermal Response}
\label{response}

The climatic relevance of the Section~\ref{flux} forcing scenarios can
be framed in terms of the ratio of $\tau_E$ to the characteristic
timescale of the forcing. For seasonal/annual variations, the natural
frequency is $\omega_{\rm orb}=2\pi/P_{\rm orb}$, whereas for
explicitly diurnal forcing one may consider $\omega_{\rm
  rot}=2\pi/P_{\rm rot}$. In this work, the fast-rotator flux maps are
diurnally averaged (Section~\ref{form}), so the dominant variability
is seasonal/annual. The slow-rotator maps, by contrast, represent a
limiting case in which strong instantaneous geometry persists over
long intervals, emphasizing the upper envelope of local forcing. A
convenient dimensionless measure of the expected damping is therefore
\begin{equation}
  \chi(\beta,\phi) \equiv \omega_{\rm orb}\,\tau_E(\beta,\phi),
  \label{eqn:chi}
\end{equation}
where $\tau_E(\beta,\phi)$ may be evaluated using a local radiating
temperature scale derived from the flux maps in Section~\ref{flux} and
an adopted effective column mass. An example of this behavior is the
seasonal lag of surface temperature relative to TOA insolation for
Earth \citep{donohoe2020a}. For Venus-like atmospheres, the radiative
environment and effective thermal inertia can be very different due to
the massive atmosphere and dominant cloud radiative effects
\citep{tomasko1980d,haus2016,limaye2018a}, so the value of $\chi$ is
expected to determine whether eccentricity- or obliquity-driven
forcing patterns in Figures \ref{fig:mod}--\ref{fig:comb} produce an
appreciable temperature modulation or are largely smoothed.

\begin{figure*}
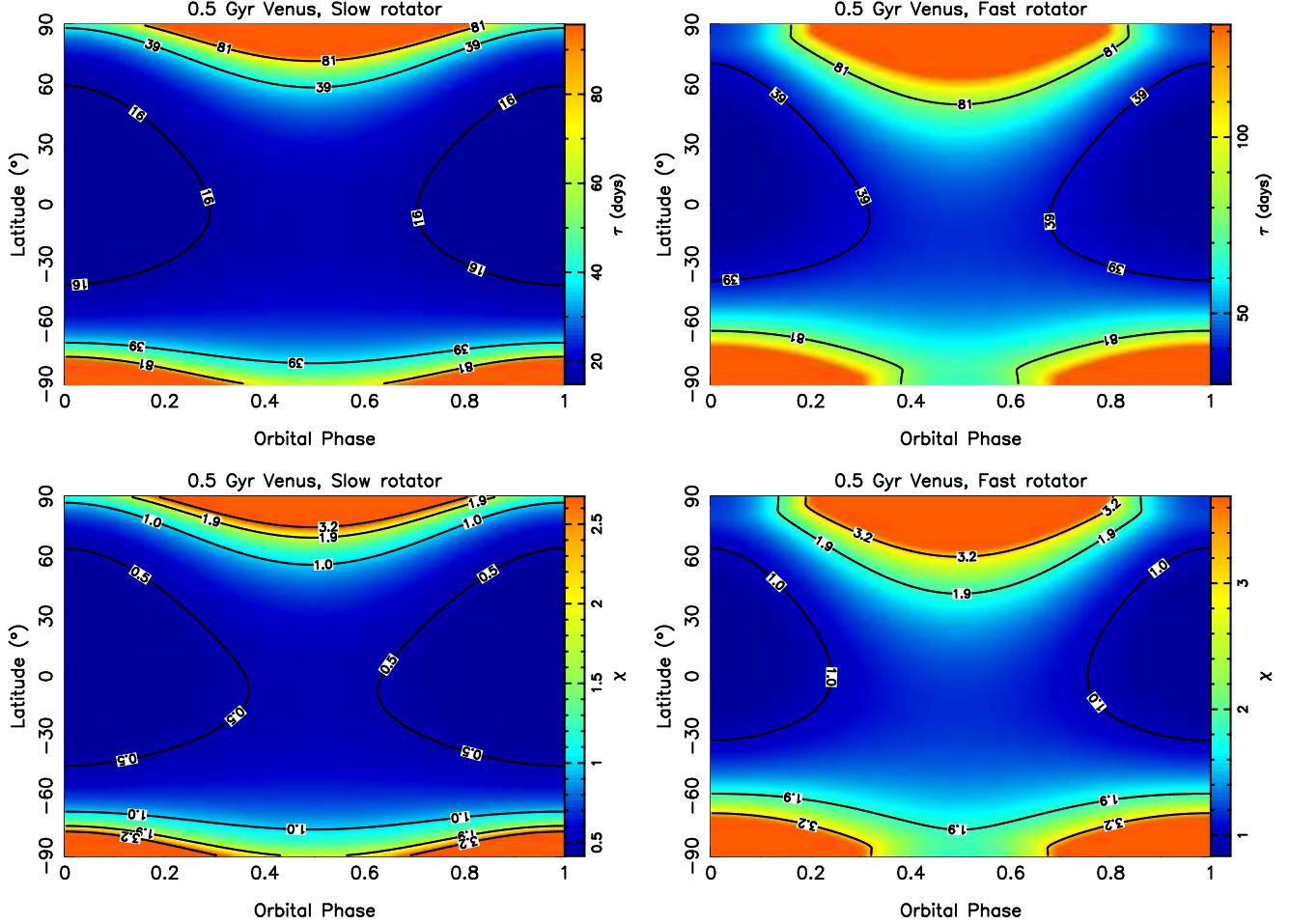

  \begin{center}
    \begin{tabular}{cc}
      \includegraphics[angle=270,width=8.5cm]{f06a.eps} &
      \includegraphics[angle=270,width=8.5cm]{f06b.eps} \\
      \includegraphics[angle=270,width=8.5cm]{f06c.eps} &
      \includegraphics[angle=270,width=8.5cm]{f06d.eps} \\
    \end{tabular}
  \end{center}
  \caption{Maps of the radiative relaxation timescale
    ($\tau_E(\beta,\phi)$; top row) and the dimensionless damping
    parameter ($\chi(\beta,\phi)$; bottom row) for the combined-factor
    case (Section~\ref{comb}), shown for the slow-rotator (left) and
    fast-rotator (right) limits. The maps are computed from
    Equation~\ref{eqn:ret} and Equation~\ref{eqn:chi} using a local
    temperature scale derived from the absorbed flux and an adopted
    effective column mass. Regions with $\chi \ll 1$ are expected to
    track seasonal forcing with minimal lag, whereas regions with
    $\chi \gg 1$ are expected to smooth variability and exhibit
    substantial phase lag. Note that the contours shown follow a
    power-law distribution to accommodate the dependency of $\tau$ and
    $\chi$ on temperature.}
  \label{fig:response}
\end{figure*}

Figure~\ref{fig:response} shows the thermal response distribution in
terms of the $\tau_E(\beta,\phi)$ (Equation~\ref{eqn:ret}; top row)
and $\chi(\beta,\phi)$ (Equation~\ref{eqn:chi}; bottom row)
calculations for the combined scenario described in
Section~\ref{comb}. These calculations assume a 1~bar CO$_2$
atmosphere, a Bond albedo of $A_B = 0.5$, an emissivity of $\epsilon =
0.9$, a specific heat capacity of $c_P = 850$~J/kg/K, and a surface
gravity of $g = 8.87$~m/s$^2$. The choice of $P_s = 1$~bar represents
a plausible early atmospheric state intermediate between a thin
primordial atmosphere and the present-day 93~bar envelope; it is
intended as a diagnostic reference pressure rather than a prediction
of early Venus surface pressure, and $T_{ao}$ is obtained from
Equation~\ref{eqn:teqigm} evaluated at each latitude--phase grid
cell. We note that the radiative relaxation timescale is
pressure-dependent, so that different atmospheric levels will exhibit
different response times \citep{showman2002}. The value adopted here
therefore characterizes the column-integrated response at the chosen
reference depth. The median values for the top-left and top-right
panels are 19.8~days and 46.1~days, respectively, demonstrating the
slower response time for rapid rotators that are more evenly
distributing the insolation flux. The median values for the
bottom-left and bottom-right panels are 0.6 and 1.3,
respectively. These values reflect the increased capability of slow
rotators in tracking seasonal forcing throughout the orbit.


\subsection{Application to the 0.5~Gyr Forcing Scenarios}
\label{apply}

Table~\ref{tab:stats} demonstrates that, at 0.5~Gyr, the mean TOA flux
is relatively insensitive to the geometric modifications explored in
Section~3, particularly for the slow-rotator cases (with mean values
$\simeq$ 1210--1230~W/m$^{2}$ across the four 0.5~Gyr rows). The
principal changes induced by obliquity and eccentricity are therefore
expressed as the redistribution of flux in latitude and orbital phase
rather than as large changes in orbit-averaged energy input. This
suggests that a single global estimate of $\tau_E$ based on a
characteristic radiating temperature (e.g., derived from the global
mean absorbed flux) will vary only weakly among the 0.5~Gyr
scenarios. In other words, the lack of explicit Equation~\ref{eqn:ret}
dependence on rotation, obliquity, and eccentricity is acceptable
since, if atmospheric properties are held fixed, the intrinsic thermal
response time is approximately constant, while the forcing differs in
pattern and temporal structure. We note, however, that an implicit
dependence exists through $T_{ao}$: changes in the geometric
parameters alter the local absorbed flux and thus the local radiating
temperature, which feeds back into $\tau_E$ via the $T_{ao}^{-3}$
scaling. This is particularly relevant when interpreting the
fast-rotator cases where the mean TOA flux changes by only a few
percent between the baseline and combined-factor scenarios
(Table~\ref{tab:stats}), so differences in expected temperature
modulation are controlled mainly by the amplitude of the seasonal
redistribution and the degree of eccentricity-driven perihelion
pulsing \citep{kane2020e,berger1978c}.

At the same time, the strong $T_{ao}^{-3}$ dependence in
Equation~\ref{eqn:ret} implies that local radiative timescales can
differ substantially within a single scenario if the flux map contains
strong maxima. This motivates computing $\tau_E(\beta,\phi)$ (or
equivalently $\chi(\beta,\phi)$) using a local temperature scale from
the flux maps. Eccentricity-driven forcing in particular produces
brief, high-flux intervals near perihelion (see bottom row of
Figure~\ref{fig:mod} and also Figure~\ref{fig:comb}), raising local
temperatures and shortening $\tau_E$ in a confined region of phase
space, even while leaving the mean flux largely unchanged. Whether
such excursions translate into meaningful atmospheric or surface
temperature modulation depends on the competition between the
perihelion-pulse duration (set by orbital geometry) and the local
radiative relaxation time (set by atmospheric thermal inertia). In the
limit $\chi\gg 1$, the response is expected to be weak and lagged,
whereas for $\chi\lesssim 1$ the atmosphere can partially track the
forcing and express a stronger seasonal signal
\citep{cronin2013b,guendelman2019,donohoe2020a}. These diagnostics
provide a direct and quantitative pathway from the Section~3 flux maps
(Figures \ref{fig:mod}--\ref{fig:comb}) and their summary statistics
(Table~\ref{tab:stats}) to expectations for the magnitude and phase of
temperature variability in early Venus scenarios. Further refinements,
including explicit dynamical heat transport and cloud feedbacks known
to be important for slowly rotating terrestrial planets, can then be
explored using GCMs as a next step
\citep{edson2011,way2016,limaye2018a}.


\section{Discussion}
\label{discussion}

This work provides a quantitative bridge between plausible early
dynamical states of Venus (rotation, obliquity, and eccentricity) and
the spatial-temporal pattern of energy input that couples to
atmospheric structure, circulation, and surface-atmosphere
exchange. The rotation endmembers primarily govern whether the forcing
resembles a sharply peaked, instantaneous geometry (slow-rotator
limit) or a smoothed, diurnally averaged seasonal pattern
(fast-rotator limit), providing an intuitive bracket on day-night
contrast and the degree of latitudinal homogenization expected from
diurnal cycling \citep{berger1978c}. An important distinction is that
the slow-rotator maps represent sub-stellar (maximum) flux at each
latitude, while the fast-rotator maps represent a true diurnal-mean
quantity. This means that the two regimes are not directly comparable
in absolute magnitude, and the energy-balance diagnostics of
Section~\ref{balance} should be understood as applying to different
spatial scales: globally representative for the fast-rotator case, but
only regionally (near the sub-stellar point) for the slow-rotator
case. Modest obliquity excursions alter the seasonal migration of the
subsolar latitude and strengthen high-latitude seasonality
\citep{barnes2016a}, while enhanced eccentricity drives strong
perihelion-centered forcing pulses, increasing peak flux and variance
without requiring a large change in the mean \citep{kane2020e}. The
interplay between eccentricity, obliquity, and the longitude of
perihelion can produce a rich variety of seasonal forcing patterns
\citep{spiegel2010b,ohno2019a,ohno2019b,guendelman2020}. These
distinct signatures suggest that early Venus could have experienced
regimes in which seasonal/annual variability was dominated either by
latitudinal redistribution (obliquity) or by temporally concentrated
energetic excursions (eccentricity), with the realized pattern
conditioned by the spin state that can plausibly arise from competing
atmospheric and solid-body tidal torques
\citep{correia2003a,correia2003b,revol2023,musseau2024}.

The climatic relevance of this redistributed forcing depends on how
insolation geometry couples to atmospheric feedbacks that are
inherently nonlinear. In particular, the response of clouds and
planetary albedo to variations in the local energy input can either
amplify or suppress seasonal contrasts, and this sensitivity is known
to be especially strong for slowly rotating terrestrial planets where
persistent dayside convection can generate optically thick cloud decks
that increase albedo and reduce surface temperatures
\citep{edson2011,barnes2016a,way2016,way2020}. For Venus, the
present-day radiative environment is itself strongly shaped by clouds
and aerosols that regulate absorbed solar energy and longwave emission
\citep{haus2016,limaye2018a}. The flux-map experiments therefore offer
a pathway to identify where and when insolation extremes occur under
plausible early dynamical states, which in turn helps to isolate which
latitudes and seasons are most likely to trigger feedbacks associated
with cloud formation, atmospheric circulation changes, and the
vertical redistribution of heating. Even if the mean energy input is
similar among cases, the enhanced spatial gradients and episodic
maxima introduced by eccentricity and rotation regimes could be
decisive for processes that respond to thresholds in temperature,
stability, or cloud microphysics rather than to the global mean alone
\citep{way2016,limaye2018a}.

Section~\ref{balance} provides a timescale-based interpretation of why
orbital and spin geometry can matter even when $\tau_E$ is not itself
an explicit function of rotation, obliquity, or eccentricity. The
radiative relaxation time is primarily set by atmospheric thermal
inertia and the characteristic radiating temperature
\citep{cronin2013b,guendelman2019}. The geometric parameters enter
through the frequency content and waveform of the forcing encoded in
$F(\beta,\phi)$, motivating the use of the dimensionless damping
parameter to determine whether seasonal/annual variability is tracked
or smoothed \citep{donohoe2020a}. In this context, eccentricity-driven
forcing introduces brief, intense perihelion excursions that may act
as energetic pulses, while obliquity-driven forcing redistributes
energy more continuously over the orbit by shifting the subsolar
latitude. Whether these differences translate into measurable
modulation of atmospheric temperatures (and thus potentially into
changes in circulation, cloud distributions, or photochemical
environments relevant to volatile loss) depends on the competition
between the forcing timescale and the local radiative adjustment time,
rather than on mean flux alone
\citep{cronin2013b,guendelman2019,kane2020e}.

A key caveat of the present analysis is that the atmospheric response
is represented using an intentionally idealized energy-balance and
gray-greenhouse framework. This is designed to provide transparent
diagnostics but necessarily omits important processes that are central
to Venus climate physics. In particular, the model does not resolve
3-D dynamics, super-rotation, wave/thermal-tide coupling, or the
microphysical and radiative feedbacks of clouds and aerosols that
control both shortwave absorption and longwave emission in the Venus
atmosphere
\citep{lebonnois2010,lebonnois2016,limaye2018a}. Furthermore, applying
a simplified radiative timescale framework beyond zero dimensions
neglects the effects of dynamical heat transport, which can
substantially modify the thermal response even in
intermediate-complexity models
\citep{showman2002,rose2017a,guendelman2022b}. The diagnostics
presented in Section~\ref{balance} should therefore be interpreted as
order-of-magnitude estimates of the expected damping regime rather
than as precise predictions of temperature amplitudes.  Accordingly,
the flux-map results should be interpreted as well-defined
boundary-condition forcing templates rather than as a complete
prediction of temperatures or circulation. The natural next step is to
apply the climate forcing scenarios within fully 3-D climate models,
such as ROCKE-3D \citep{way2017b} and the LMD Venus GCM
\citep{lebonnois2010,lebonnois2016}, to quantify how rotation regime,
obliquity, and eccentricity jointly shape the coupled response of
dynamics, clouds, and radiative transfer.  Such experiments would
allow the insolation extremes and seasonal redistributions identified
here to be tested against self-consistent cloud formation, heat
transport, and vertically resolved radiative cooling, improving
constraints on the sensitivity of early Venus climate states and
potential transition pathways \citep{edson2011,way2016}.

The results from the provided flux maps may provide useful context for
the expected data that will be produced from the next generation of
Venus missions. In particular, the latitude-orbital phase flux maps
and idealized energy-balance framework identify where and when the TOA
forcing is maximized and how strongly it is expected to be smoothed by
atmospheric opacity. While the present framework cannot
self-consistently predict cloud-driven albedo variability,
thermal-tide forcing, or convective stability, the forcing maps
identify the orbital phases and latitudes at which such processes are
most likely to be energetically driven. These patterns can inform the
vertical stability structure that the NASA Deep Atmosphere Venus
Investigation of Noble gases, Chemistry, and Imaging (DAVINCI) mission
will probe during descent through the Venusian atmosphere
\citep{garvin2022}. Likewise, by connecting energy balance scenarios
to candidate climate and resurfacing pathways, our results help frame
the timing and environmental conditions of major geologic transitions
that the NASA Venus Emissivity, Radio Science, InSAR, Topography, and
Spectroscopy (VERITAS) mission will test via global radar
interferometric topography and gravity field constraints on
lithospheric structure and geologic evolution
\citep{cascioli2021}. Finally, the combined orbital forcing cases
provide boundary-condition guidance for coupled climate-geology
interpretations relevant to the ESA EnVision mission, whose
synergistic radar and spectroscopic investigations are designed to
link atmospheric state, surface properties, and recent or ongoing
geologic activity \citep{widemann2023}. Our maps help identify
energetically favored regimes for interpreting observed spatial
patterns in surface emissivity, apparent alteration, and atmospheric
variability.


\section{Conclusions}
\label{conclusions}

In this work we quantified how plausible early dynamical states of
Venus modify the distribution of TOA stellar forcing by constructing
latitude-orbital phase flux maps for both the present epoch and an
early epoch at 0.5~Gyr. We explored endmember rotation regimes
(slow-rotator and fast-rotator limits) and a set of maximum obliquity
and eccentricity cases motivated by recent dynamical studies, as well
as a combined scenario that represents an upper envelope of
seasonal/annual forcing variability. The resulting maps demonstrate
that rotation primarily controls the instantaneous versus diurnally
averaged character of the forcing, setting the amplitude and sharpness
of local flux maxima, while obliquity and eccentricity imprint
distinct seasonal signatures. Because the slow-rotator maps represent
sub-stellar (maximum) forcing and the fast-rotator maps represent
diurnally averaged forcing, the two representations bracket the range
of local forcing regimes rather than providing directly comparable
global quantities.  For example, obliquity drives latitudinal
migration of the subsolar forcing and strengthens high-latitude
seasonality, whereas eccentricity produces temporally concentrated
perihelion pulses that increase peak flux and variance. Across the
explored parameter space, the mean incident flux changes relatively
little compared to the redistribution of energy in latitude and
orbital phase, indicating that early Venus insolation variability is
governed more by when and where energy is delivered than by large
changes in the orbit-averaged energy input.

To connect these forcing patterns to climate-relevant behavior, we
used an idealized atmospheric energy-balance framework to define a
radiative relaxation timescale that characterizes the thermal
adjustment of an atmospheric column. The analysis highlights that the
intrinsic relaxation time depends primarily on atmospheric thermal
inertia and the characteristic radiating temperature rather than
directly on orbital and rotational dynamical components. Instead, the
dynamical parameters influence climate response through the frequency
content and waveform of the forcing encoded in the flux maps. This
timescale-based interpretation provides a physical criterion for when
insolation variability is likely to be expressed as temperature
modulation versus being smoothed into an approximately steady mean
state. Together, the flux maps and idealized response metrics
establish a set of dynamical forcing templates that can be used to
guide and interpret future 3-D climate simulations of early Venus and
to connect orbital-spin evolution scenarios to the observable
atmospheric and surface signatures targeted by upcoming Venus
missions. The acquired data from these missions will create a further
diagnostic dataset from which climate forcing scenarios will paint a
more reliable picture of the evolutionary history for our sibling
world.


\section*{Acknowledgements}

The author wishes to thank the two anonymous reviewers, whose feedback
helped to improve the manuscript. This research has made use of the
Habitable Zone Gallery at hzgallery.org. The results reported herein
benefited from collaborations and/or information exchange within
NASA's Nexus for Exoplanet System Science (NExSS) research
coordination network sponsored by NASA's Science Mission Directorate.


\section*{Author Contributions}

S.R.K. performed the analysis, prepared all figures, and wrote the
manuscript text.


\section*{Funding}

The author declares that no funds, grants, or other support were
received during the preparation of this manuscript.


\section*{Competing Interests}

The author has no relevant financial or non-financial interests to
disclose.


\section*{Ethics Declaration}

Not applicable.




\end{document}